\documentclass[a4paper,preprint,aps,pra,showkeys,superscriptaddress]{revtex4-2}
\usepackage{physics}
\usepackage{lineno,hyperref,soul,amsmath,amssymb,color,graphicx,bm}

\usepackage{graphicx}
\usepackage[caption=false]{subfig}   
\usepackage{xcolor}       
\usepackage{longtable}
\usepackage{xparse}
\NewExpandableDocumentCommand\mcc{O{1}m}
{\multicolumn{#1}{l}{#2}}
\usepackage[utf8]{inputenc} 
\usepackage[english]{babel}
\usepackage{enumitem, etoolbox, tabularx, makecell, booktabs, float, nccmath}
\newcolumntype{P}[1]{>{\abovedisplayskip=\abovedisplayshortskip \belowdisplayskip=\belowdisplayshortskip}p{#1}}
\usepackage{booktabs,siunitx,makecell}
\usepackage{dcolumn}
\usepackage[colorinlistoftodos]{todonotes}
\usepackage{marginnote}
\usepackage{fancyhdr}
\usepackage{adjustbox}
\usepackage{multirow}
\usepackage[normalem]{ulem}
\usepackage{xcolor}
\usepackage{lipsum}

\makeatletter
\usepackage{miller}
\newcommand\footnoteref[1]{\protected@xdef\@thefnmark{\ref{#1}}\@footnotemark}
\makeatother
 
\reversemarginpar

\hypersetup{
	pdfnewwindow=false,      
	colorlinks=true,       
	linkcolor=red,          
	citecolor=blue,        
	filecolor=magenta,      
	urlcolor=cyan           
}
\usepackage{siunitx}
\usepackage{xfrac}
\usepackage{accents}
\numberwithin{equation}{section}
\makeatletter
\def\tagform@#1{\maketag@@@{(#1)\@@italiccorr}}
\makeatother
\let\orgautoref\autoref

\let\orgautoref\autoref

\let\orgautoref\autoref

\renewcommand{\autoref}[1]
{%
	\def\equationautorefname{Eq.}%
	\def\figureautorefname{Fig.}%
	\def\paperautorefname{paper}%
	\def\subfigureautorefname{Fig.}%
	\orgautoref{#1}%
}

\renewcommand{\autoref}[1]
{%
	\def\equationautorefname{Eq.}%
	\def\figureautorefname{Fig.}%
	\def\paperautorefname{paper}%
	\def\subfigureautorefname{Fig.}%
	\orgautoref{#1}%
}

\makeatletter
\def\tagform@#1{\maketag@@@{\ignorespaces#1\unskip\@@italiccorr}}%
\let\orgtheequation\theequation%
\def\theequation{(\orgtheequation)}%
\makeatother%
\usepackage{soul}
\usepackage{longtable}
\usepackage{enumitem}
\usepackage{etoolbox}
\makeatletter
\patchcmd\frontmatter@PACS@format{\addvspace{11\p@}}{}{}{}
\pretocmd\frontmatter@keys@format{\addvspace{11\p@}}{}{}
\patchcmd{\titleblock@produce}
  {\@pacs@produce\@pacs\@keywords@produce\@keywords}
  {\@keywords@produce\@keywords\@pacs@produce\@pacs}
  {}{}
\makeatother
\newlength\mylength
\usepackage[caption=false]{subfig}
\usepackage{chemformula}

\begin{document}
\title{Evaluating Moment Tensor Potential in Ag–Cu Alloy: Accuracy, Transferability, and Phase Diagram Fidelity}
\author{Mashroor S. Nitol}
\email{mash@lanl.gov}
 \affiliation{Center for Integrated Nanotechnologies, Los Alamos National Laboratory, Los Alamos, NM 87544, USA}
  \author{Marco J. Echeverría Iriarte}
\affiliation{CEA DAM DIF, Bruyères-le-Châtel, 91297 Arpajon, France}
\affiliation{Université Paris-Saclay, LMCE 91280 Bruyères-le-Châtel, France}

\author{Doyl E. Dickel}
\affiliation{Michael W. Hall School of Mechanical Engineering, Mississippi State University, Starkville, MS 39762, USA}

\author{Saryu J. Fensin}
 \affiliation{Center for Integrated Nanotechnologies, Los Alamos National Laboratory, Los Alamos, NM 87544, USA}

\begin{abstract}
A Moment Tensor Potential (MTP) has been developed for the Cu–Ag binary alloy and its accuracy, transferability, and thermodynamic fidelity evaluated. The model was trained on a diverse dataset encompassing solid, liquid, and interfacial configurations derived from density functional theory (DFT) calculations. Benchmarking against experiment and DFT data demonstrated significant improvements over the widely used classical Embedded Atom Method (EAM) potential, particularly in predicting defect energetics, surface properties, and the eutectic phase diagram. Despite a slight underestimation of Ag’s melting point, the MTP model achieved consistent accuracy across elemental and binary systems without direct fitting to high-temperature phase transitions. The predicted eutectic temperature and composition were found in close agreement with experimental observations. These results establish MTP as a robust framework for modeling immiscible metallic systems and pave the way for its integration into large-scale atomistic simulations where both fidelity and generalizability are essential.
\end{abstract}

\keywords{Ag, Cu, transition materials, phase diagram, molecular dynamics, machine learning, interatomic potential}
\maketitle 
\newpage
\section{Introduction}
The copper–silver (Cu–Ag) alloy system has been the focus of sustained interest due to its unique combination of immiscibility in the solid state and eutectic formation in the liquid phase\cite{sakai1997ultra,massalski1986binary,balluffi1978vacancy}. This system has been widely studied for its applications in electronics, catalysis, and soldering, where phase behavior and surface diffusion critically influence material performance\cite{massalski1986binary}. The relatively large lattice mismatch between Cu and Ag, approximately 12\%, has rendered this system an archetype for exploring heteroepitaxial growth, interface energetics, and atomic transport phenomena\cite{morgenstern2004direct,ozolicnvs1998effects}. As a result, the Cu–Ag binary alloy has served as a benchmark for testing interatomic potential models and simulation methodologies across length scales.

To model the interatomic interactions in the Cu–Ag system, the Embedded Atom Method (EAM)\cite{daw1984embedded,foiles1986embedded} has been widely employed, including a parameterization specifically constructed to reproduce the thermodynamic and structural properties of both elements and their binary mixtures\cite{mishin2005interatomic}. Using a physically motivated functional form and a limited number of fitting parameters, a potential using the EAM formalism has been shown to capture essential features such as cohesive energies, lattice constants, and elastic moduli with moderate success\cite{williams2006embedded}. However, its transferability to surface and defect-related properties remains limited. As demonstrated by \citet{wu2009cu}, the EAM potential significantly underestimates surface and stacking fault energies, particularly in low-coordination environments where directional bonding becomes important. To address these deficiencies, the surface EAM (SEAM)\citet{wu2009cu} was introduced by incorporating angular corrections into the EAM framework, thereby improving surface energetics while retaining the computational efficiency of EAM. Nevertheless, the improved surface accuracy came at the cost of reparameterization and limited generalizability across complex alloy configurations, underscoring the inherent limitations of fixed functional forms.

In recent years, machine learning interatomic potentials (ML-IAPs) have emerged as a compelling alternative capable of bridging the accuracy of density functional theory (DFT) with the scalability of classical potentials\cite{vita2024spline,morawietz2021machine,zuo2020performance,deringer2019machine,behler2016perspective,kobayashi2017neural,dickel2021lammps}. In ML-IAPs, the form of the potential energy surface is not assumed a priori but is instead inferred from data using local atomic environment descriptors that are invariant to rotation, translation, and permutation\cite{mishin2021machine}. Methods such as the Moment Tensor Potential (MTP)\cite{shapeev2016moment}, Gaussian Approximation Potential (GAP)\cite{bartok2013representing}, and high-dimensional neural network potentials (NNPs)\cite{behler2016perspective} have been developed to this end. Among them, MTP has been recognized for its favorable trade-off between accuracy and computational cost\cite{zuo2020performance}. The MTP framework constructs rotationally covariant tensor descriptors and systematically contracts them into scalar invariants, allowing flexible and complete representations of atomic interactions\cite{gubaev2019accelerating,podryabinkin2019accelerating,novikov2018automated,nitol2025moment}. 

In this study, a MTP has been constructed for the Cu–Ag binary system with the goal of accurately capturing thermodynamic properties across the solid and liquid phases. Special emphasis has been placed on the reproduction of the binary phase diagram, which exhibits a eutectic point at approximately 1052~K and a composition near 28.1~at.\% Cu\cite{massalski1986binary}. The phase diagram belongs to a simple eutectic type with complete immiscibility in the solid state and limited solubility in the liquid, making it a stringent benchmark for assessing interatomic potentials. The EAM model\cite{williams2006embedded}, has been able to qualitatively reproduce the overall shape of the phase diagram but deviates in the predicted eutectic temperature and composition due to insufficient resolution of subtle entropic and enthalpic contributions near the solid–liquid coexistence region. Accurate prediction of the eutectic point requires a potential that simultaneously captures the energetics of disordered liquid structures and the delicate free energy differences between solid phases, which has traditionally posed a challenge for fixed-form empirical models.

By contrast, the MTP developed in this work has been trained on a diverse dataset incorporating solid, liquid, and interfacial configurations, enabling accurate interpolation across phase boundaries. The predicted eutectic point is found to be in close agreement with experimental data, reflecting the MTP's ability to resolve fine features in the free energy landscape. In addition, improvements are observed in key defect and surface properties, including the generalized stacking fault energy (GSFE) of Ag and the surface energies of both Cu and Ag. Although the melting point of Ag is slightly underestimated by approximately $\sim$100K, the MTP consistently outperforms the EAM reference model across multiple property classes. These findings support the use of machine-learned interatomic potentials as versatile tools for high-fidelity modeling of phase stability and interfacial phenomena in binary metallic systems.

\section{Methodology}
\subsection{Training database}
A training dataset with controlled variation was constructed to represent the relatively simple atomic environments characteristic of Cu–Ag alloys. First-principles calculations were performed using the Vienna \textit{Ab initio} Simulation Package (VASP)~\cite{hafner2008ab}, version 6.3.2. The Perdew–Burke–Ernzerhof (PBE) exchange-correlation functional within the generalized gradient approximation (GGA)~\cite{perdew1996generalized} was employed. The Brillouin zone was sampled using a Monkhorst–Pack $k$-point mesh with a spacing finer than $2\pi \times 0.01$~\AA$^{-1}$. The plane-wave basis was defined with a kinetic energy cutoff of 520~eV. Gaussian smearing with a width of 0.2~eV was applied to facilitate Brillouin zone integration. Electronic self-consistency was achieved when the total energy difference between iterations fell below $10^{-6}$~eV, and ionic relaxation was performed until the maximum force on any atom was less than 0.01~eV/\AA.

Following prior work~\cite{nitol2025moment,nitol2023hybrid}, the dataset includes strained configurations of FCC Cu and Ag with lattice distortions up to 15\% in order to accurately reproduce elastic constants. Additional high-symmetry polymorphs—including BCC, HCP, $A15$, $\beta$-Sn, simple cubic (SC), and diamond cubic (DC)—were incorporated to ensure that the correct ground-state structures are robustly favored by the potential. For pure elemental systems, supercells of various sizes were constructed, incorporating random atomic displacements up to 0.5~\AA{} and isotropic box size perturbations of $\pm$5\% to mimic thermal fluctuations and volumetric expansion. Relaxed configurations of surface slabs, generalized stacking faults (GSFEs), monovacancies, and interstitial defects were also included to improve defect transferability.

For binary structures, ordered intermetallic compounds were included despite the experimental immiscibility of Cu and Ag in the solid phase, in order to prevent the potential from artificially stabilizing unphysical ground states. Solid solution configurations were generated by introducing Cu atoms into Ag lattices and vice versa at various concentrations, with atomic perturbations and box strains applied analogously to those in the pure-element datasets. This approach ensures thermodynamic consistency and robustness across the compositional and structural landscape of the Cu–Ag system.

The final Ag–Cu MLIP training database comprises a diverse and well-balanced set of atomic configurations designed to enable robust potential fitting across both elemental and alloyed environments. It includes pure Ag and Cu structures, as well as Ag–Cu alloys, spanning a range of crystallographic types and physical phenomena. Key categories include thermally perturbed FCC and BCC/HCP supercells, sheared unit cells, gamma-surface (GSFE) configurations, isolated atoms, free surfaces, and point defects. Notably, over 2,400 configurations are dedicated to perturbed FCC alloys to ensure accurate interpolation across concentration ranges. The dataset covers both relaxed and thermally excited atomic environments, providing a foundation for accurately modeling energy, force, and stress behavior under varied thermomechanical conditions.

The dataset is publicly available at the authors’ \href{https://gitlab.com/mtp_potentials/agcu/-/tree/main/00.training_database?ref_type=heads}{GitLab repository}. A supporting Python \href{https://gitlab.com/mtp_potentials/tialv/-/tree/main/Training?ref_type=heads}{wrapper} is also provided for converting the database into formats compatible with other transferable training frameworks.

\begin{figure*}[!htbp]
\noindent \centering
\includegraphics[width=\textwidth]{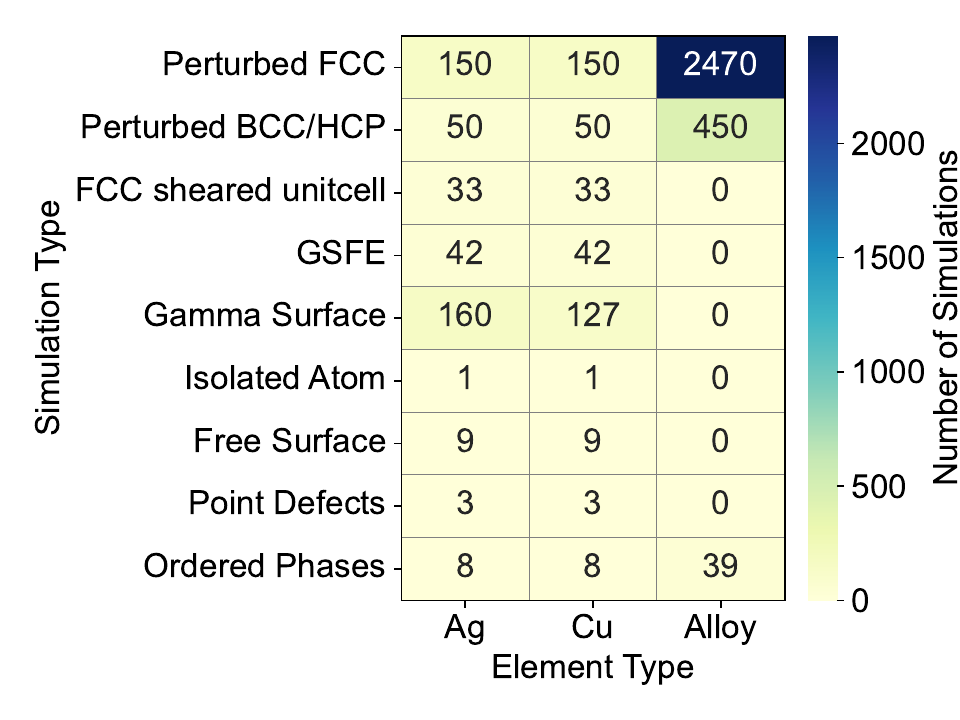}
\caption{Visualization of the Cu–Ag MLIP training database composition across simulation types and elemental content.}
\label{fig:database}
\end{figure*}

\subsection{Moment tensor potential training}
The MTP framework has been successfully applied to a broad range of material systems, including elemental metals and multicomponent alloys~\cite{gubaev2019accelerating, podryabinkin2019accelerating}. Within this formalism, the local atomic environment is described using moment tensors $M_{\mu,\nu}$, defined as:
\begin{equation}
    M_{\mu,\nu}(\mathbf{n}_i) = \sum f_\mu(|\mathbf{r}_{ij}|, z_i, z_j) \, \mathbf{r}_{ij} \otimes \ldots \otimes \mathbf{r}_{ij},
\end{equation}
where $\mathbf{n}_i$ characterizes the chemical identity and spatial arrangement of neighboring atoms around the $i^{\text{th}}$ atom. The vector $\mathbf{r}_{ij}$ denotes the relative position from atom $i$ to atom $j$, while $z_i$ and $z_j$ indicate their atomic types. The radial dependence is captured through the basis functions $f_\mu$, and angular information is encoded via the rank-$\nu$ tensor product of $\mathbf{r}_{ij}$. The accuracy and computational efficiency of the potential are governed by two key parameters: the cutoff radius ($r_{\text{cut}}$), which defines the interaction range, and the maximum expansion level ($\text{lev}_{\text{max}}$), which controls the number of basis tensors and thus the expressiveness of the model. The latter is typically set as an even integer between 2 and 28.

Following recent work by ~\citet{nitol2025moment}, a systematic grid search was performed to optimize the MTP hyperparameters. Expansion levels ranging from 16 to 20 and cutoff radii from 5.0~\AA to 6.0~\AA (in 0.1~\AA increments) were explored. For each parameter set, 20 independent MTP models were trained with randomized initial weights, resulting in a total of 600 trained models. From this ensemble, the 50 models with the lowest root mean squared error (RMSE) in energy per atom were selected for detailed evaluation. These models were benchmarked against key physical properties, including elastic constants, GSFE, and melting points of pure elements.

The optimal model was identified at $\text{lev}_{\text{max}} = 16$ with a cutoff radius of $r_{\text{cut}} = 5.9$~\AA. This configuration yielded accurate predictions for elastic constants, GSFE, and melting points across both Cu and Ag. The final MTP exhibited an energy RMSE of 0.008eV/atom, a force RMSE of 0.018eV/\AA, and a stress RMSE of 0.127GPa. During training, the loss function weights were set to $w_{e} = 1$ for energy, $w_{f} = 0.01$ for forces, and $w_{s} = 0.001$ for stress components. This weighting scheme was designed to prioritize energy accuracy while ensuring that force and stress information contributed meaningfully to the optimization process.

A 90:10 train–validation split was used to facilitate model convergence and prevent overfitting. All training and model assessments were conducted using the MLIP package~\cite{novikov2020mlip}. Molecular dynamics simulations were performed in LAMMPS~\cite{thompson2022lammps}, and atomic structure visualization and trajectory analysis were carried out using OVITO~\cite{stukowski2009visualization}, which was also used to generate figures presented in this work.

\begin{figure*}[!htbp]
\noindent \centering
\includegraphics[width=0.8\textwidth]{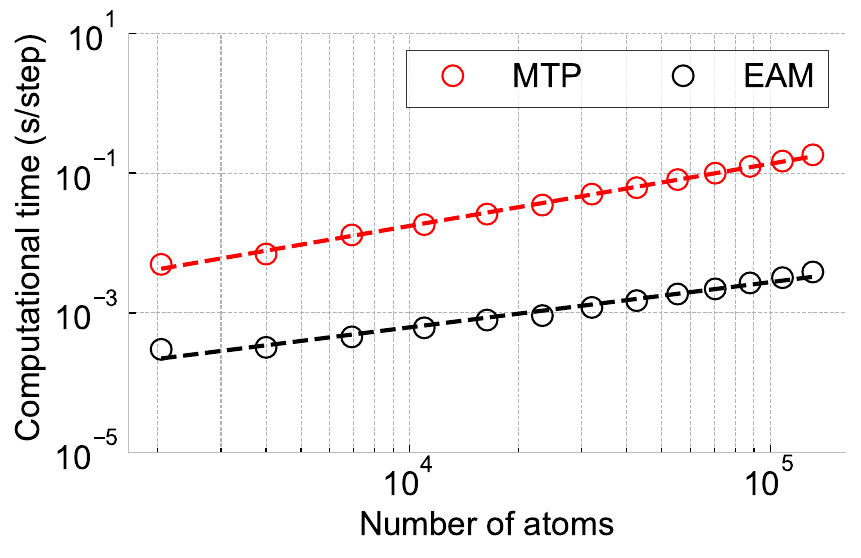}
\caption{
Log--log plot of computational time per MD step (NPT ensemble at 300~K) versus number of atoms for MTP (red) and EAM (black) in a Cu--25\%Ag system. Each data point represents a 5,000-step simulation performed on a single node with 128 processor cores. Dashed lines indicate linear fits, confirming near-linear scaling for both models.
}
\label{fig:eam_mtp_speed}
\end{figure*}

To assess the scalability of the MTP relative to a classical potential, benchmark simulations were conducted to compare computational performance against the EAM potential by \citet{williams2006embedded} (for the remainder of the text, it will be referred to as EAM). ~\autoref{fig:eam_mtp_speed} presents a log--log plot of the average computational time per molecular dynamics (MD) step as a function of the number of atoms, ranging from 2,048 to 131,072 atoms of Cu-25\%Ag. Each simulation was conducted on a single compute node comprising 128 processor cores. Each timing corresponds to a fixed trajectory length of 5,000 MD steps. The MTP results are shown in red, and the EAM results are shown in black, with linear fits overlaid as dashed lines to highlight scaling behavior.

At the smallest system size (2,048 atoms), the EAM model achieves a time per step of approximately 0.00051 s/step, while the MTP model requires around 0.00496 s/step, indicating a performance overhead of nearly 10x. This relative cost remains consistent across all tested system sizes. For the largest system benchmarked (131,072 atoms), EAM simulations proceed at approximately 0.0183 s/step, whereas MTP requires 0.1834 s/step — again showing a consistent order-of-magnitude increase in time per step.

Despite this disparity in absolute timing, both models exhibit linear scaling behavior on the log–log plot, with fitted slopes close to unity. This observation confirms that the computational cost grows proportionally with system size, suggesting that the parallelization and data locality of both implementations remain efficient even at large scales. Importantly, the consistent scaling slope indicates that the MTP's additional computational cost arises from its descriptor evaluation and basis expansion complexity, rather than from poor algorithmic scaling.

Overall, while MTP incurs a 10x computational penalty relative to EAM, its linear scaling and accuracy gains make it a viable choice for medium- to large-scale simulations where high fidelity is required, especially in modeling properties like stacking faults, surfaces, and phase transitions that are not captured at DFT accuracy by classical potentials.

\section{Results}
\subsection{Force validation}
Forces were evaluated by differentiating the total energy with respect to atomic positions, as defined by
\[
\vec{f}_{i} = -\frac{\partial E}{\partial \vec{r}_{i}},
\]
and compared directly against \textit{ab initio} forces obtained from DFT calculations. The MTP model was trained using energy, force, and stress components, enabling the construction of an efficient yet physically meaningful training set. The resulting force RMSE values were 0.017~eV/\AA{} for pure Ag, 0.022~eV/\AA{} for pure Cu, and 0.014~eV/\AA{} for the Ag--Cu alloy system. These values demonstrate excellent agreement between MTP predictions and DFT reference data across both unary and binary configurations.

\begin{figure}[htbp]
    \centering
    \includegraphics[width=0.9\linewidth]{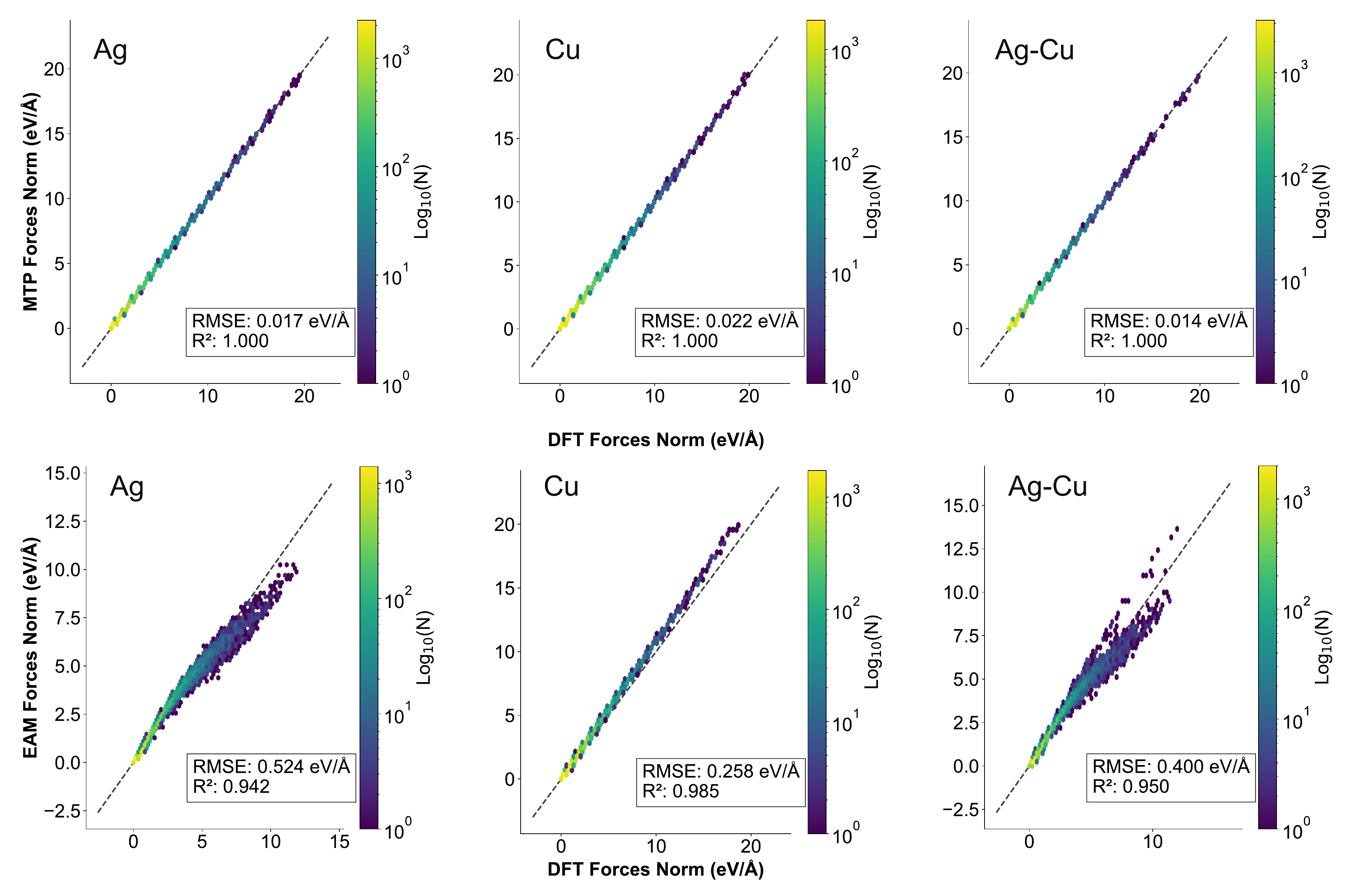}
    \caption{Comparison of MTP-(top row) and EAM- (bottom row) predicted force norms with DFT reference values for pure Ag, pure Cu, and the Ag--Cu alloy. Each point corresponds to the magnitude of force on an atom in a given configuration.}
    \label{fig:all_force}
\end{figure}

\autoref{fig:all_force} presents scatter plots comparing the norms of predicted and DFT-calculated atomic forces for each system. The alignment of data points along the diagonal reflects the high accuracy of the MTP in reproducing force magnitudes across a wide range of local environments. This validation confirms the model's suitability for dynamic simulations, where accurate force prediction is critical for capturing processes such as diffusion, defect migration, and thermally activated transitions. Achieving low force RMSEs in both elemental and alloy systems provides a solid foundation for reliable atomistic modeling within the broader Cu--Ag compositional space.

\subsection{Cold curve}
\begin{figure*}[!htbp]
\noindent \centering
\includegraphics[width=0.8\textwidth]{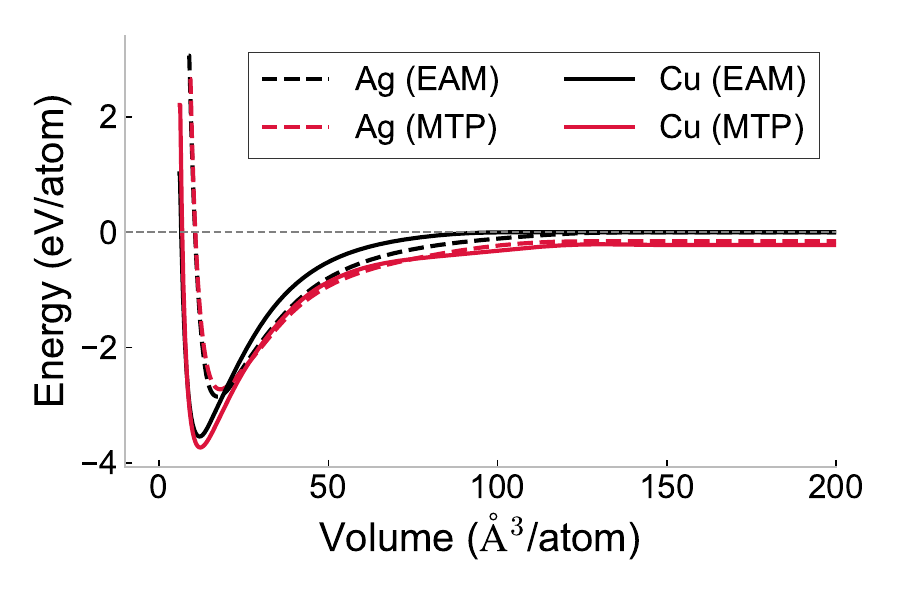}
\caption{
Cold curves for Cu and Ag predicted by MTP and EAM. While both agree near equilibrium, MTP shows physically consistent energy rise under compression despite lacking training data below 85\% volume, highlighting its robustness in extrapolation. MTP predicted energies for Ag and Cu are offset by approximately 0.15 and 0.22 eV/atom, respectively, relative to the zero-energy reference. This discrepancy arises because, in DFT, the total energy of an isolated atom is not set to zero. The MTP inherits this reference offset as it attempts to match the DFT isolated-atom energies.}
\label{fig:ev}
\end{figure*}

To evaluate the accuracy and physical reliability of different interatomic potentials, we computed and compared the cold curves—total energy as a function of atomic volume—for elemental Cu and Ag using the MTP and EAM potentials. These curves are shown in \autoref{fig:ev}, which highlights key differences in behavior across a range of volumes, including those corresponding to substantial compressive strains.

The EAM potential, as expected, yield smooth parabolic-like energy–volume (E–V) curves with minima located near the equilibrium atomic volumes for both Cu and Ag. The behavior is consistent with classical potential models, which are typically designed to maintain physical plausibility across a relatively broad range of thermodynamic states\cite{mishin2021machine}.

The MTP predictions demonstrate not only excellent agreement with EAM near equilibrium volumes, but also exhibit superior physical fidelity under volumetric compression. Notably, despite being trained only up to 15\% compression, the MTP cold curves do not display unphysical behavior outside the training domain. Instead, they show a monotonically increasing energy profile under compression, consistent with the expected rise in short-range repulsive forces as atomic distances decrease.

This is in stark contrast with some neural network potentials (NNPs) reported in the literature\cite{nitol2022unraveling}, which have occasionally demonstrated non-physical behavior, such as decreasing energy under high compression, when extrapolated beyond their training set. Such behavior arises from the over-flexibility and lack of physical priors in purely mathematical regression models.

The observed robustness of the MTP potential can be attributed to its systematic polynomial basis, which, while flexible, retains structural constraints that help preserve physical trends such as short-range repulsion and energy convexity. This makes MTP a more predictively stable and physically informed choice for simulations involving moderate to high compressive strains.

\subsection{Basic properties}

\autoref{table:cu_ag_comparison} summarizes a range of material properties for Ag and Cu as predicted by the EAM and MTP potentials, alongside experimental measurements and DFT reference data. The EAM potential was primarily fitted to experimental observations, whereas the MTP potential was trained on DFT-calculated datasets. Despite its relatively simple analytical form and small number of fitting parameters, EAM demonstrates impressive accuracy in predicting cohesive energy, lattice parameters, elastic constants, and point defect formation energies—particularly for Cu. However, EAM exhibits notable deviations in properties that are highly sensitive to electronic structure, such as surface energies and stacking fault energies, especially in the case of Ag. In contrast, MTP shows significantly better agreement with DFT across these quantities, including $\gamma_{usf}$, $\gamma_{isf}$, and surface energies for various low-index crystallographic planes.

For structural energy differences between competing crystal phases (e.g., FCC $\rightarrow$ BCC, HCP, SC, etc.), both EAM and MTP show good correspondence with DFT, suggesting reasonable transferability across phase space. Overall, while EAM offers efficient and accurate predictions for bulk and point-defect properties, MTP achieves superior fidelity to DFT, particularly for surface- and fault-related energetics, albeit with higher computational cost. Quantitatively, the RMSE computed across all available properties with respect to DFT is 46.01 for EAM and 15.99 for MTP.

\begin{table}
\centering
\caption{Comparison of material properties of Ag and Cu predicted by EAM and MTP, alongside experimental and DFT values. Cohesive energy ($E_{coh}$) is reported in eV; lattice constants ($a$) in \AA; elastic constants ($C_{ij}$) in GPa; structural energy differences ($\Delta E$) in meV; and point defect energies—including vacancy formation energy ($E_{vac}$), vacancy migration energy ($E_{m}$), octahedral ($E_{octa}$), and tetrahedral ($E_{tetra}$) interstitials—are in eV. The vacancy formation volume ($\Omega_v$) is given relative to the equilibrium atomic volume ($\Omega_0$). Surface energies ($E_{surf}$) and stacking fault energies ($\gamma_{sf}$) are presented in mJ/m$^2$. Experimental and DFT references are provided in the footnotes; unless otherwise noted, DFT data are from the present study.}
\footnotesize
\resizebox{\textwidth}{!}{%
\begin{tabular*} {\textwidth}{@{\extracolsep{\fill}} lcccc|cccc}
\toprule
& \multicolumn{4}{c|}{{Ag}} & \multicolumn{4}{c}{{ Cu}} \\ \cline{2-5} \cline{6-9}
Property &  Expt & DFT & EAM & MTP &   Expt & DFT & EAM & MTP \\
\midrule
$E_{coh}$  & 2.96$^{(c)}$ & 2.60 & 2.85 & 2.57   & 3.49$^{(c)}$ & 3.61 & 3.54 & 3.57 \\
$a$  & 4.086$^{(a)}$ & 4.15 & 4.09 & 4.15  & 3.62$^{(a)}$ & 3.63 & 3.61 & 3.63 \\
$C_{11}$  & 122$^{(a)}$ & 100.5$^{(b)}$,113.55 & 124.24 & 112.90    & 169$^{(a)}$ & 174.96$^{(b)}$,178.99 & 169.89 & 178.99 \\
$C_{12}$  & 92$^{(a)}$ & 85.25$^{(b)}$,84.24 & 93.87 & 81.03   & 122$^{(a)}$ & 121.57$^{(b)}$,123.64 & 122.61 & 123.64 \\
$C_{44}$ & 45.5$^{(a)}$ & 39.16$^{(b)}$,41.34 & 46.42 & 44.12   & 75.3$^{(a)}$ & 76.45$^{(b)}$,79.55 & 76.19 & 79.55 \\
$E_{vac}$  & 1.1$^{(f)}$ & 1.03$^{(g)}$,0.73 & 1.10 & 0.78  & 1.27$^{(d)}$ & 1.09$^{(e)}$,1.06 & 1.27 & 0.98 \\
$\Omega_v/\Omega_0$ &  0.94$^{(i)}$ & -- & 0.68 & 0.69  &  0.78$^{(i)}$ & -- & 0.70 & 0.72 \\
$E_{m}$ & 0.66$^{(f)}$ & -- & 0.655 & 0.625 & 0.71$^{(f)}$ & -- & 0.689 & 0.754 \\
$E_{tetra}$  & -- & 3.89$^{(g)}$,3.44 & 4.11 & 3.46  &  -- & 3.72$^{(e)}$,3.78 & 3.60 & 3.74 \\
$E_{octa}$  & -- & 2.38$^{(g)}$,2.95 & 3.49 & 2.89  &  -- & 3.33$^{(e)}$,3.40 & 3.18 & 3.32 \\
$E_{surf}(100)$  &-- & 820$^{(g)}$ & 940 & 858 &   -- & 1486$^{(e)}$ & 1345 & 1437 \\
$E_{surf}(110)$  & -- & 870$^{(g)}$ & 1017 & 898  &  -- & 1571$^{(e)}$ & 1475 & 1517 \\
$E_{surf}(111)$  & -- & 760$^{(g)}$ & 862 & 739  & -- & 1329$^{(e)}$ & 1239 & 1283 \\
$\gamma_{usf}$\hkl<11-2>(111)  & -- & 92.01$^{(b)}$,100.4$^{(h)}$,91.54 & 118.24 & 97.88   & -- & 160.52$^{(b)}$,163.7$^{(h)}$,154.99 & 163.26 & 165.94 \\
$\gamma_{isf}$\hkl<11-2>(111)  & -- & 14.49$^{(b)}$,17.8$^{(h)}$,17.26 & 14.63 & 19.48   & -- & 41.83$^{(b)}$,38.5$^{(h)}$,41.69 & 44.01 & 40.43 \\
$E_{\text{FCC}\rightarrow\text{BCC}}$  & -- & 0.027 & 0.033 & 0.033  & -- & 0.032 & 0.034 & 0.036 \\
$E_{\text{FCC}\rightarrow\text{HCP}}$ & -- & 0.003 & 0.004 & 0.005 &  -- & 0.008 & 0.007 & 0.008 \\
$E_{\text{FCC}\rightarrow\text{SC}}$  & -- & 0.326 & 0.327 & 0.307 &  -- & 0.458 & 0.448 & 0.459 \\
$E_{\text{FCC}\rightarrow\omega}$  & -- & 0.060 & 0.064 & 0.086 &  -- & 0.072 & 0.077 & 0.077 \\
$E_{\text{FCC}\rightarrow\text{A15}}$ & -- & 0.091 & 0.087 & 0.063 &  -- & 0.097 & 0.087 & 0.086 \\
$E_{\text{FCC}\rightarrow\text{DC}}$  & -- & 0.777 & 0.798 & 0.798 & -- & 1.021 & 0.989 & 1.017 \\
$E_{\text{FCC}\rightarrow\beta\text{Sn}}$  & -- & 0.237 & 0.237 & 0.224 &  -- & 0.319 & 0.314 & 0.311 \\
\bottomrule
\end{tabular*}
}
\vspace{2mm}
\begin{minipage}{0.95\textwidth}
\footnotesize
$^{(a)}$Expt.\cite{warlimont2018springer}, 
$^{(b)}$DFT\cite{su2019density}, 
$^{(c)}$Expt.\cite{kittel2018introduction}, 
$^{(d)}$Expt.\cite{siegel1978vacancy}, 
$^{(e)}$DFT\cite{liyanage2024machine}, 
$^{(f)}$DFT\cite{balluffi1978vacancy}, 
$^{(g)}$DFT\cite{andolina2021robust}, 
$^{(h)}$DFT\cite{hunter2013dependence}, 
$^{(i)}$Expt.\cite{williams2006embedded}
\end{minipage}
\label{table:cu_ag_comparison}

\end{table}
\subsection{Phonon dispersion}
\begin{figure*}[!htbp]
\noindent \centering
\includegraphics[width=\textwidth]{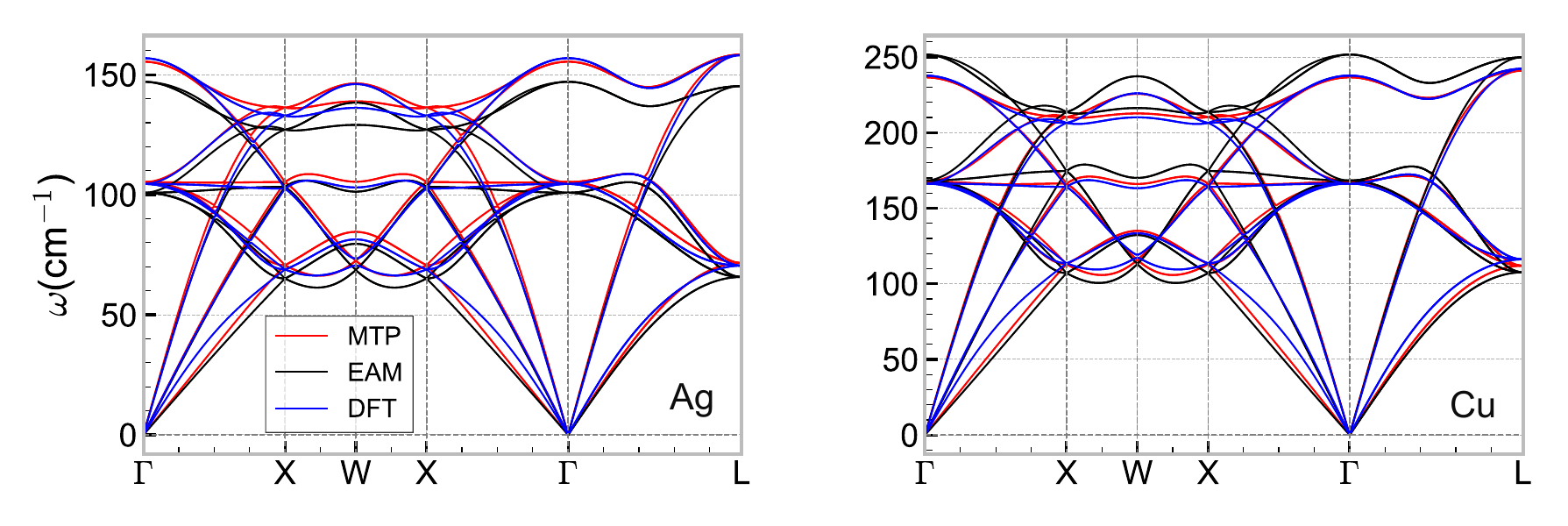}
\caption{Phonon dispersion relations for Cu and Ag predicted by EAM and MTP potentials, shown alongside DFT reference data.}

\label{fig:phonon_dispersion}
\end{figure*}

Phonon dispersion curves provide critical insight into the vibrational and thermodynamic behavior of crystalline materials. In this study, phonon dispersion relations for Cu and Ag were computed using the EAM and MTP potentials and compared against DFT reference data. The results are presented in \autoref{fig:phonon_dispersion}.

The MTP potential shows excellent agreement with DFT across all high-symmetry directions in the Brillouin zone. Both the longitudinal and transverse acoustic branches are accurately captured, especially near the $\Gamma$-point, where the slopes of the acoustic modes correspond to elastic constants. Furthermore, MTP accurately reproduces the curvature and energy levels of zone-boundary phonons.

In contrast, the EAM potential, though able to reproduce the overall shape of the dispersion relations, exhibits minor  discrepancies. The transverse acoustic branches are generally softened near the zone boundaries, particularly in Ag, suggesting a systematic underestimation of shear stiffness. Additionally, deviations in the longitudinal acoustic branch and the flattening of higher-frequency modes indicate limited transferability of EAM to vibrational properties. 

To quantitatively assess the accuracy of the potentials, RMSE were computed between the predicted phonon frequencies and the DFT reference data. For Ag, an RMSE of 0.0458cm$^{-1}$ was obtained using MTP, which is less than half the RMSE of 0.1050cm$^{-1}$ obtained using EAM. A similar trend was observed for Cu, where RMSE values of 0.0624cm$^{-1}$ and 0.1624cm$^{-1}$ were found for MTP and EAM, respectively. The reduced RMSE values across both elements suggest that MTP provides a more accurate representation of the harmonic force constants and exhibits improved transferability for modeling vibrational and thermodynamic properties.

\subsection{Stacking fault and gamma surface}

In order to evaluate the accuracy and transferability of MTP for Ag and Cu, its predictions of generalized stacking fault energies on the FCC \hkl{111} plane were compared to DFT and EAM potentials. The key stacking fault energies considered include the unstable stacking fault energy ($\gamma_{\mathrm{USF}}$), intrinsic stacking fault energy ($\gamma_{\mathrm{ISF}}$), the aligned stacking fault energy ($\gamma_{\mathrm{ASF}}$), and the peak GSFE value along the $\langle110\rangle$ direction ($\gamma_{p110}$). These values are derived from energy surfaces sampled along the \hkl<112> and $\langle110\rangle$ crystallographic directions.

\begin{figure*}[!htbp]
    \centering
    \includegraphics[width=\linewidth]{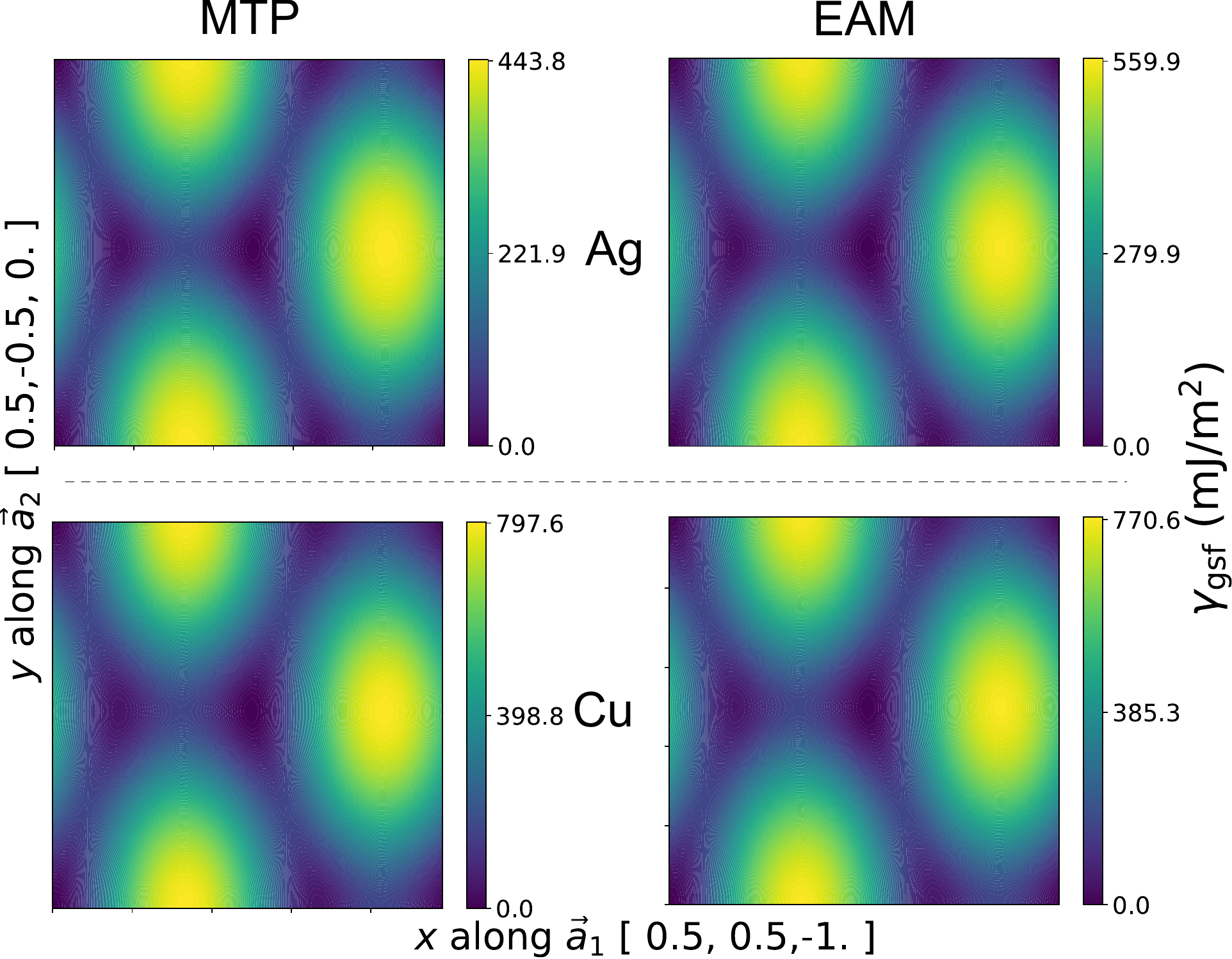}
    \caption{2D GSFE surface plots on the FCC \{111\} plane for Ag and Cu using MTP (left) and EAM (right). Displacements are along $\langle112\rangle$ and $\langle110\rangle$ directions.}
    \label{fig:gsfe_surface}
\end{figure*}

\begin{figure*}[htbp]
    \centering
    \includegraphics[width=\linewidth]{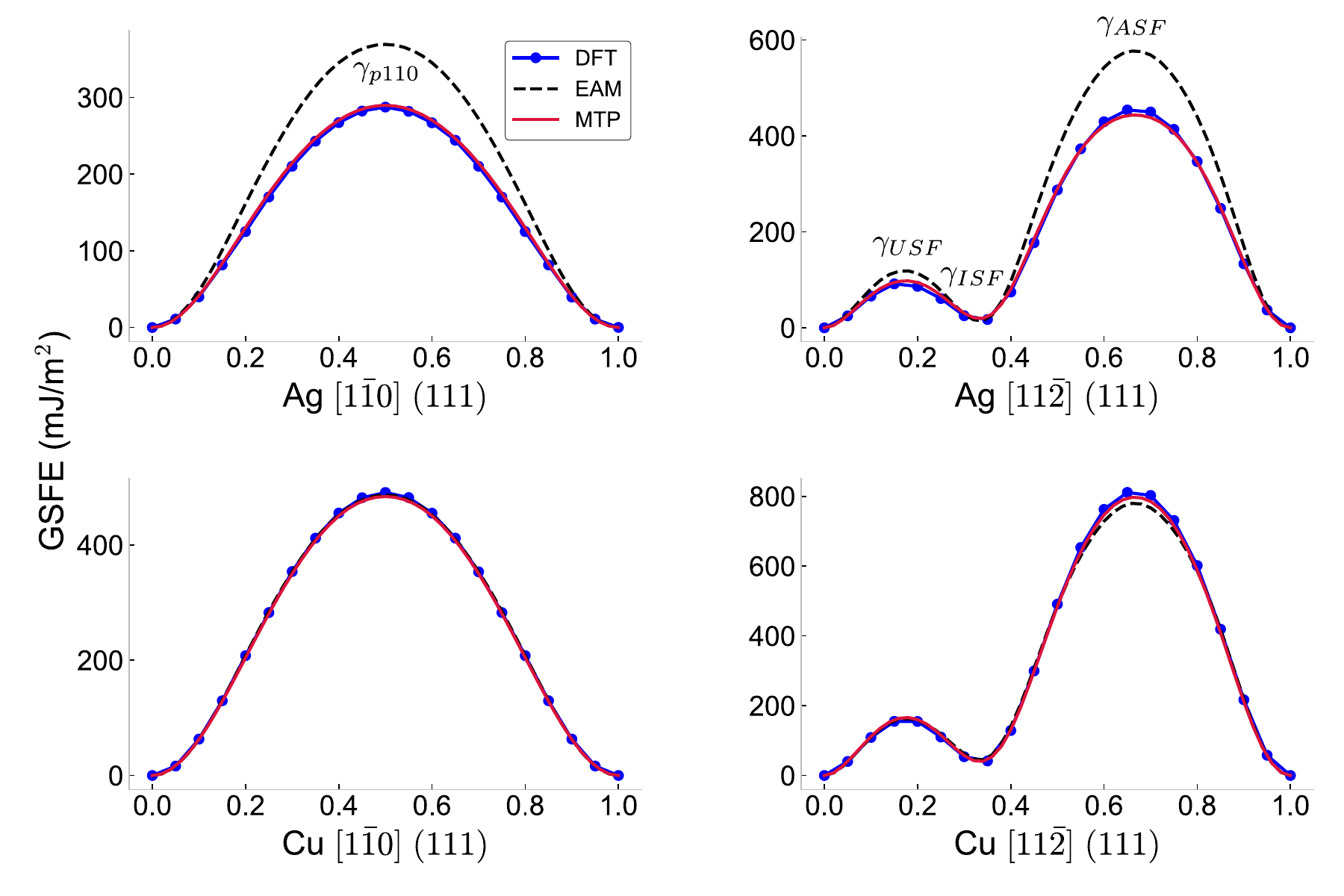}
    \caption{$\gamma$ line of GSFE energy curves for Ag and Cu along the $\langle112\rangle$ and $\langle110\rangle$ directions on the \{111\} plane. MTP and EAM predictions are compared with DFT benchmarks.}
    \label{fig:gsfe_1d}
\end{figure*}

As shown in ~\autoref{fig:gsfe_surface}, the MTP reproduces the qualitative and quantitative features of the DFT-calculated GSFE surface for both Ag and Cu. The overall energy landscape—including the location of energy maxima and minima—is captured far more accurately by MTP than by the EAM potential, which tends to exaggerate fault energies and smooth out critical features. \autoref{fig:gsfe_1d} further illustrates the $\gamma$-line GSFE curves along both \hkl<112> and \hkl<110> directions. Quantitatively, the MTP model predicts $\gamma_{\mathrm{USF}}$, $\gamma_{\mathrm{ISF}}$, and $\gamma_{\mathrm{ASF}}$ within 2–8\% of the DFT values for both elements. In contrast, the EAM model exhibits discrepancies of up to 30\%, particularly in $\gamma_{\mathrm{ASF}}$ and $\gamma_{p110}$ for Ag.

These findings echo the conclusions of \citet{su2019density}, who emphasized that the shape and accuracy of the GSFE surface critically affect dislocation core dissociation, stacking fault widths, and overall plastic deformation mechanisms in FCC metals. Their work demonstrated that Cu and Ag closely follow ideal FCC behavior, with displacements to $\gamma_{\mathrm{USF}}$ and $\gamma_{\mathrm{ISF}}$ aligning well with hard-sphere expectations. The quantitative values are summarized in ~\autoref{tab:gsfe_values}, highlighting the agreement of MTP and EAM predictions with DFT.

\begin{table*}[htbp]
\centering
\caption{Comparison of GSFE values (in mJ/m$^2$) for Ag and Cu along the $\langle112\rangle$ and $\langle110\rangle$ directions on the \{111\} plane as predicted by DFT, MTP, and EAM.}
\label{tab:gsfe_values}
\begin{tabular}{lcccccccc}
\toprule
\multirow{2}{*}{Method} & \multicolumn{4}{c}{Ag} & \multicolumn{4}{c}{Cu} \\
\cmidrule(lr){2-5} \cmidrule(lr){6-9}
 & $\gamma_{\mathrm{USF}}$ & $\gamma_{\mathrm{ISF}}$ & $\gamma_{\mathrm{ASF}}$ & $\gamma_{p110}$ & $\gamma_{\mathrm{USF}}$ & $\gamma_{\mathrm{ISF}}$ & $\gamma_{\mathrm{ASF}}$ & $\gamma_{p110}$ \\
\midrule
DFT & 91.54 & 17.26 & 454.11 & 287.48 & 154.99 & 41.69 & 811.16 & 491.16 \\
MTP & 97.88 & 19.48 & 443.43 & 289.79 & 165.94 & 40.43 & 797.03 & 484.16 \\
EAM & 118.60 & 14.75 & 576.94 & 369.29 & 163.07 & 44.85 & 779.60 & 486.77 \\
\bottomrule
\end{tabular}
\end{table*}

In this context, current MTP potential achieves DFT-level fidelity while retaining the computational efficiency needed for large-scale simulations. which makes it a robust tool for studying dislocation dynamics, twinning, and deformation in FCC systems, as guided by the DFT-informed insights of \citet{su2019density}.

\subsection{Thermal expansion and melting point}
\begin{figure*}[htbp]
    \centering
    \subfloat{\includegraphics[width=0.5\linewidth]{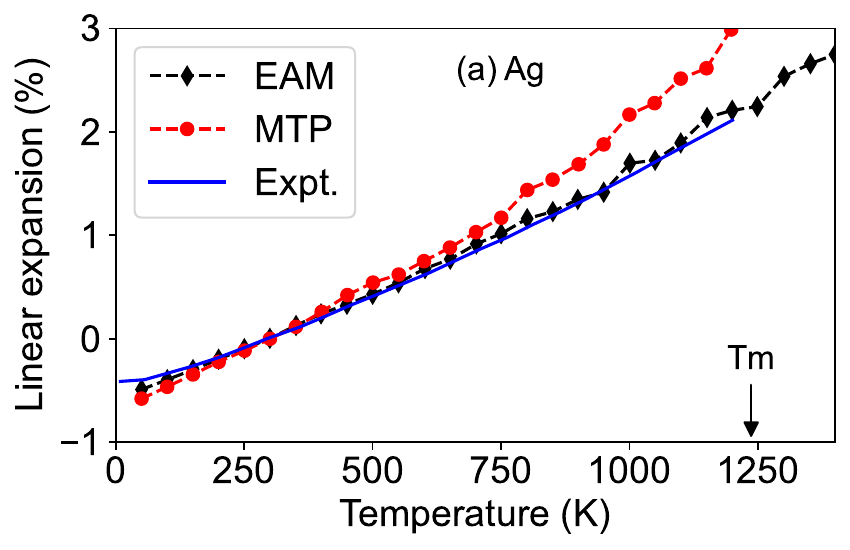}}
    \subfloat{\includegraphics[width=0.5\linewidth]{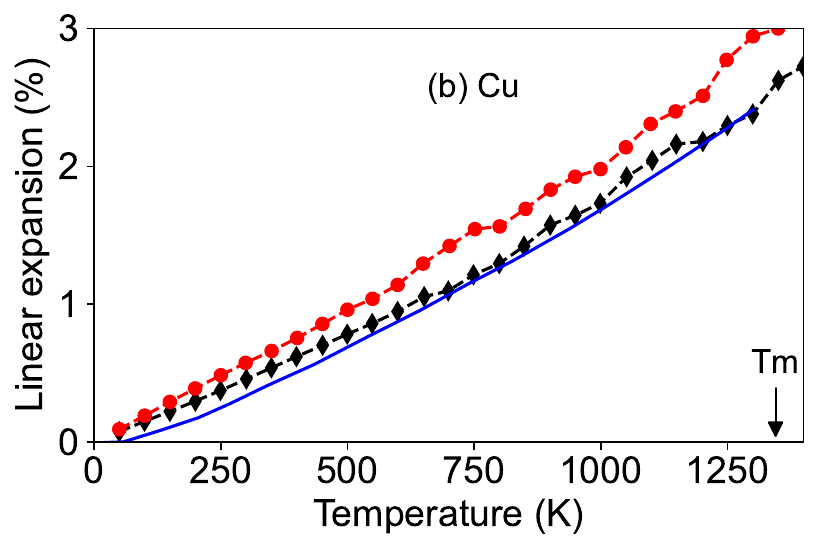}}
    \caption{Linear thermal expansion of (a) Ag (relative to 300 K) and (b) Cu (relative to 0 K), computed using the NPT ensemble with the EAM potential. Experimental reference values are relative to different temperatures based on the available data sources, taken from \cite{touloukian1975thermal}. Arrows indicate the experimental melting points $T_m$}
    \label{fig:thermal}
\end{figure*}

Thermal expansion behavior of FCC Ag and Cu was computed using a molecular dynamics approach in which atoms in a 4000-atom cubic supercell (10$\times$10$\times$10 unit cells, each containing 4 atoms) were allowed to move around their equilibrium positions, while the simulation volume was allowed to fluctuate under a zero-pressure condition (NPT ensemble). As shown in \autoref{fig:thermal}, the results demonstrate excellent agreement with experimental data \cite{touloukian1975thermal}.

The melting temperature $T_m$ was computed using a 64,000-atom supercell containing a solid–liquid interface, following the methodology described in \cite{nitol2023hybrid}. While \textit{ab initio} predictions for Cu \cite{baty2021cu} and Ag \cite{baty2021ag} tend to slightly underestimate the melting temperature, the EAM potential—explicitly fitted to reproduce experimental melting points—achieves high accuracy. In contrast, the MTP potential was trained only on thermally perturbed atomic configurations without direct fitting to melting data. Nevertheless, MTP yields melting points in good agreement with \textit{ab initio} results. For Ag, the MTP-predicted melting point is approximately 100 K lower than the experimental value, indicating that MTP is reproducing the physics captured by \textit{ab initio} calculations rather than aligning precisely with experiment.
\begin{figure*}[htbp]
    \centering
    \includegraphics[width=\linewidth]{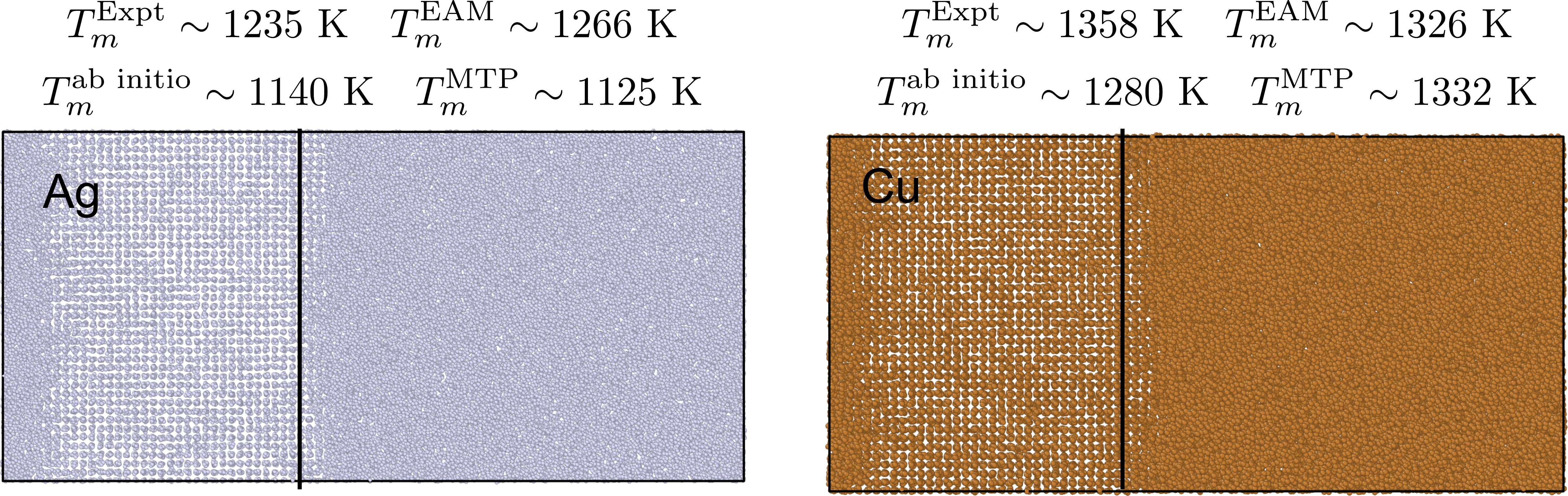}
    \caption{Melting points of pure elements computed using the MTP potential. Experimental, \textit{ab initio}, and EAM reference values are taken from \cite{touloukian1975thermal}, \cite{baty2021cu}, \cite{baty2021ag}, and \cite{williams2006embedded}, respectively.}
    \label{fig:pure_melt}
\end{figure*}

\subsection{Binary phase diagram}
Accurately modeling the Ag–Cu system across varying concentrations is essential for capturing its complex thermodynamic behavior. A reliable interatomic potential must not only reproduce key material properties but also predict phase transitions as a function of temperature and composition. The MTP successfully captures the phase behavior of the Ag–Cu alloy up to high Ag/Cu concentrations, as modeled using the semi-grand-canonical Monte Carlo (SGCMC) method~\cite{nitol2025moment} implemented in LAMMPS.

In SGCMC simulations, the system is initialized with pure Cu or Ag equilibrated at a specified temperature and chemical potential (\( \mu \)). The LAMMPS SGCMC fix employs the Metropolis acceptance criterion~\cite{sadigh2012scalable}, which includes a chemical potential difference term, $\Delta\mu$. At each molecular dynamics timestep, ten attempts are made to swap a Cu atom with an Ag atom, with swaps accepted according to the Metropolis algorithm~\cite{metropolis1953equation}. Each simulation involves 2048 atoms and runs for 100 ps. The same procedure is followed for Ag-rich and Cu-rich phases.

The composition at which the system transitions from a Cu-rich phase to an Ag-rich phase is identified as the forward transition point. Conversely, as Ag concentration is gradually reduced from an Ag-rich phase, the point where the system reverts back to a Cu-rich phase marks the reverse transition. This composition defines the maximum Ag content at which the Ag-rich phase remains stable before crossing back into the Cu-rich regime.

The phase transitions predicted by the MTP model show strong agreement with experimental data. Simulations were conducted at temperatures of 800~K, 850~K, and 900~K. \autoref{tial pd} presents the resulting hysteresis loops. The Cu-rich to Ag-rich phase boundaries are observed at Ag concentrations of 1.73\%, 2.54\%, and 3.19\% at 800~K, 850~K, and 900~K, respectively. The reverse Ag-rich to Cu-rich boundaries are found at 95.80\%, 94.78\%, and 93.89\%, respectively, indicating a narrow composition window for phase coexistence.

\begin{figure*}[!htbp]
    \centering
    \includegraphics[width=\textwidth]{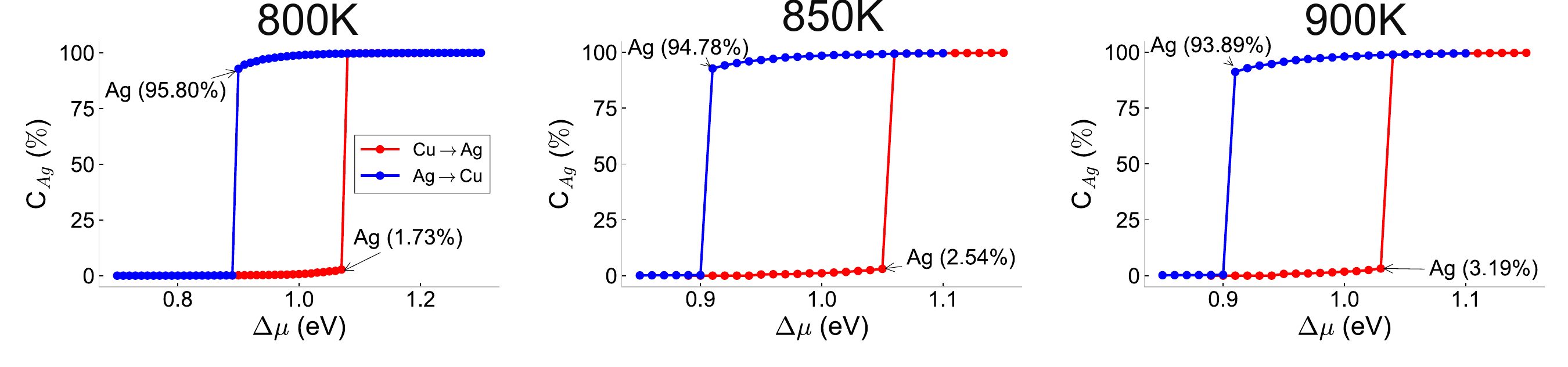}
    \caption{Hysteresis loops obtained from SGCMC simulations of the Ag–Cu alloy at 800~K, 850~K, and 900~K, indicating phase transitions between Cu-rich and Ag-rich phases.}
    \label{tial pd}
\end{figure*}

\begin{figure*}[!htbp]
    \centering
    \includegraphics[width=\textwidth]{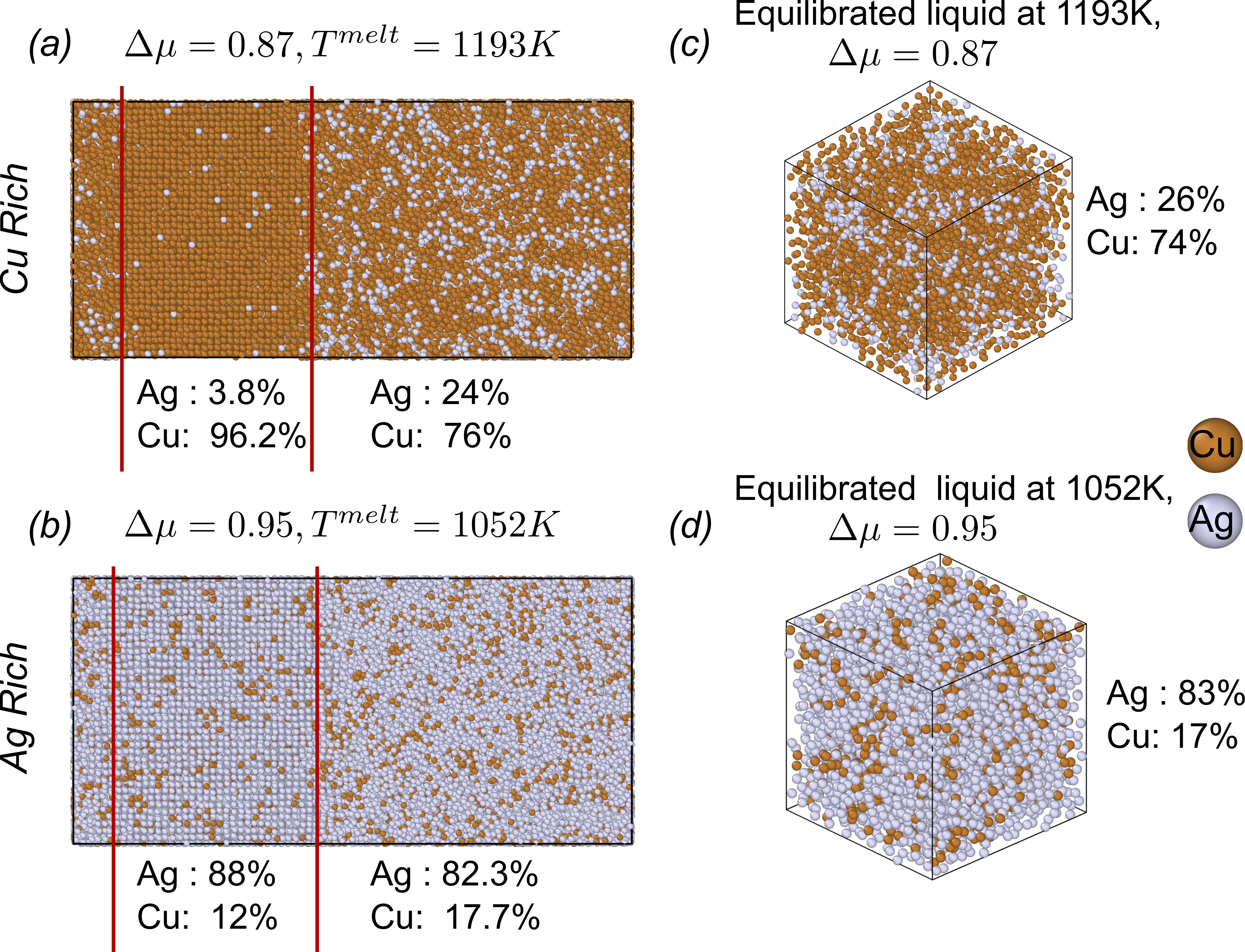}
    \caption{Snapshots from MC simulations used to determine the solidus and liquidus lines for Ag–Cu. (a, b) Interface evolution as a function of $\Delta\mu$ at fixed temperature. (c, d) Equilibrated alloy compositions corresponding to solidus and liquidus points at the melting temperature.}
    \label{pd snap}
\end{figure*}

The simulation setup begins with a two-phase configuration consisting of solid and liquid regions of the alloy, prepared similarly to the pure element melting simulations. A planar solid–liquid interface was equilibrated at a fixed temperature while systematically varying the chemical potential difference ($\Delta\mu$) between Cu and Ag atoms. As shown in \autoref{pd snap}(a,b), the direction of interface motion was monitored to identify the equilibrium $\Delta\mu$ value corresponding to zero net movement—indicating thermodynamic coexistence. For each $\Delta\mu$ increment, 50,000 MC steps per atom were performed, and the equilibrium point was determined within an uncertainty of $\pm$0.01 eV/atom. Once the equilibrium $\Delta\mu$ was obtained at a given temperature, a separate MC simulation was run under this condition to sample the stable compositions of the coexisting solid and liquid phases. These simulations, also spanning 50,000 MC steps, yielded the composition of the solidus and liquidus phases at each temperature, as shown in \autoref{pd snap}(c,d). This method ensures a consistent and accurate determination of phase boundaries across alloy compositions.

After performing simulations at various temperatures and compositions across the phase diagram, the eutectic behavior of the Ag–Cu system was re-evaluated using the developed MTP. The results demonstrate a significant improvement in the prediction of the eutectic point compared to EAM potential. Previously EAM estimated the eutectic temperature to be approximately 935~K, with a eutectic composition of 0.46 (atomic fraction of Ag), and solubility limits for Cu in Ag and Ag in Cu of 0.03 and 0.937, respectively. These values deviate notably from experimental measurements, which place the eutectic temperature at 1053~K, the eutectic composition at 0.601, and the solubility limits at 0.049 and 0.860~\cite{touloukian1975thermal}.

By contrast, the current MTP model was specifically trained to more accurately capture thermodynamic properties relevant to high-temperature behavior, including eutectic transitions. Using this model, the eutectic temperature was found to be 1000.84~K, with a eutectic composition of 0.57. The solubility limits were determined to be 0.04 for Cu in Ag and 0.901 for Ag in Cu. While the eutectic temperature predicted by MTP remains slightly below the experimental value, it represents a substantial improvement over the EAM prediction and shows strong agreement with values obtained from \textit{ab initio} methods, which are known to slightly underestimate melting points of noble metals such as Ag.

\begin{figure*}[!htbp]
    \centering
    \includegraphics[width=\textwidth]{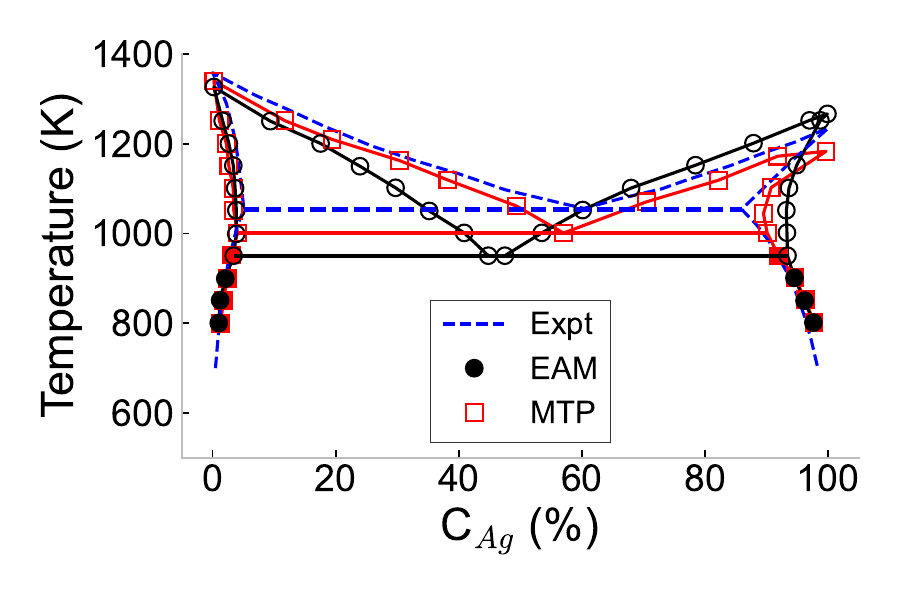}
    \caption{Calculated and experimental~\cite{touloukian1975thermal} phase diagram of the Cu–Ag system. Filled markers represent compositions obtained from SGMC simulations in the solid phase region. Open markers denote values derived from solid–liquid interface simulations used to determine melting behavior.}
    \label{tial pd}
\end{figure*}

\section{Conclusion}

The present study demonstrates that MLIPs, specifically the MTP, can bridge the longstanding trade-off between accuracy and efficiency in modeling phase behavior and defect properties of binary metallic systems. Through systematic training on a DFT-informed dataset encompassing bulk, defected, and interfacial environments, the MTP model was shown to significantly outperform the classical EAM potential, especially in capturing energetics sensitive to local structural heterogeneity such as stacking faults and surfaces.
A key strength of the MTP lies in its robustness across configurations it was not explicitly trained on. Despite the absence of direct fitting to phase diagram data or melting behavior, the predicted eutectic temperature and composition closely match experimental benchmarks. This underscores the extrapolative reliability of the MTP’s descriptor-based formalism, which systematically encodes atomic interactions while preserving physical symmetries. Unlike neural-network-based MLIPs, MTP does not suffer from instability under high compression or extrapolative breakdown, as demonstrated by its accurate cold curves and phonon spectra.
Importantly, the inclusion of pure-element and binary configurations in the training process did not compromise the fidelity of unary properties. This suggests that the MTP framework can be systematically extended to multicomponent alloys without degrading baseline elemental accuracy—a limitation often observed in overfitted empirical models. With a training protocol that incorporates physical priors and thermodynamic diversity, the model generalizes well across phase space and retains transferability from elemental systems to concentrated alloys.
The demonstrated ability to model eutectic behavior, solid–liquid transitions, and stacking fault energetics with DFT-level fidelity establishes MTP as a scalable and physically grounded approach for next-generation alloy simulations. Future extensions to ternary or quaternary systems appear feasible, provided that training sets are similarly comprehensive. Thus, MTP offers a viable path toward reliable, extensible, and thermodynamically consistent MLIPs for metallic systems.

\section*{Data availability}
The potential, training database, and sample calculations are available on the author's GitLab page:
\url{https://gitlab.com/mtp_potentials/agcu}.

\section*{Supplemental information}
The supplemental information presents the convex hull analysis of Ag–Cu alloys and the atomistic dislocation core structures of Ag and Cu.

\clearpage
\bibliographystyle{IEEEtranN} 
\bibliography{AgCu}

\begin{thebibliography}{42}
\providecommand{\natexlab}[1]{#1}
\providecommand{\url}[1]{#1}
\csname url@samestyle\endcsname
\providecommand{\newblock}{\relax}
\providecommand{\bibinfo}[2]{#2}
\providecommand{\BIBentrySTDinterwordspacing}{\spaceskip=0pt\relax}
\providecommand{\BIBentryALTinterwordstretchfactor}{4}
\providecommand{\BIBentryALTinterwordspacing}{\spaceskip=\fontdimen2\font plus
\BIBentryALTinterwordstretchfactor\fontdimen3\font minus
  \fontdimen4\font\relax}
\providecommand{\BIBforeignlanguage}[2]{{%
\expandafter\ifx\csname l@#1\endcsname\relax
\typeout{** WARNING: IEEEtranN.bst: No hyphenation pattern has been}%
\typeout{** loaded for the language `#1'. Using the pattern for}%
\typeout{** the default language instead.}%
\else
\language=\csname l@#1\endcsname
\fi
#2}}
\providecommand{\BIBdecl}{\relax}
\BIBdecl

\bibitem[Sakai and Schneider-Muntau(1997)]{sakai1997ultra}
Y.~Sakai and H.-J. Schneider-Muntau, ``Ultra-high strength, high conductivity
  cu-ag alloy wires,'' \emph{Acta Materialia}, vol.~45, no.~3, pp. 1017--1023,
  1997.

\bibitem[Massalski et~al.(1986)Massalski, Okamoto, Subramanian, Kacprzak, and
  Scott]{massalski1986binary}
T.~B. Massalski, H.~Okamoto, P.~Subramanian, L.~Kacprzak, and W.~W. Scott,
  \emph{Binary alloy phase diagrams}.\hskip 1em plus 0.5em minus 0.4em\relax
  American society for metals Metals Park, OH, 1986, vol.~1, no.~2.

\bibitem[Balluffi(1978)]{balluffi1978vacancy}
R.~Balluffi, ``Vacancy defect mobilities and binding energies obtained from
  annealing studies,'' \emph{Journal of nuclear materials}, vol.~69, pp.
  240--263, 1978.

\bibitem[Morgenstern et~al.(2004)Morgenstern, Braun, and
  Rieder]{morgenstern2004direct}
K.~Morgenstern, K.-F. Braun, and K.-H. Rieder, ``Direct imaging of cu dimer
  formation, motion, and interaction with cu atoms on ag (111),''
  \emph{Physical review letters}, vol.~93, no.~5, p. 056102, 2004.

\bibitem[Ozoli{\c{n}}{\v{s}} et~al.(1998)Ozoli{\c{n}}{\v{s}}, Wolverton, and
  Zunger]{ozolicnvs1998effects}
V.~Ozoli{\c{n}}{\v{s}}, C.~Wolverton, and A.~Zunger, ``Effects of anharmonic
  strain on the phase stability of epitaxial films and superlattices:
  Applications to noble metals,'' \emph{Physical Review B}, vol.~57, no.~8, p.
  4816, 1998.

\bibitem[Daw and Baskes(1984)]{daw1984embedded}
M.~S. Daw and M.~I. Baskes, ``Embedded-atom method: Derivation and application
  to impurities, surfaces, and other defects in metals,'' \emph{Physical Review
  B}, vol.~29, no.~12, p. 6443, 1984.

\bibitem[Foiles et~al.(1986)Foiles, Baskes, and Daw]{foiles1986embedded}
S.~Foiles, M.~Baskes, and M.~S. Daw, ``Embedded-atom-method functions for the
  fcc metals cu, ag, au, ni, pd, pt, and their alloys,'' \emph{Physical review
  B}, vol.~33, no.~12, p. 7983, 1986.

\bibitem[Mishin(2005)]{mishin2005interatomic}
Y.~Mishin, ``Interatomic potentials for metals,'' \emph{Handbook of Materials
  Modeling: Methods}, pp. 459--478, 2005.

\bibitem[Williams et~al.(2006)Williams, Mishin, and
  Hamilton]{williams2006embedded}
P.~Williams, Y.~Mishin, and J.~Hamilton, ``An embedded-atom potential for the
  cu--ag system,'' \emph{Modelling and Simulation in Materials Science and
  Engineering}, vol.~14, no.~5, p. 817, 2006.

\bibitem[Wu and Trinkle(2009)]{wu2009cu}
H.~H. Wu and D.~R. Trinkle, ``Cu/ag eam potential optimized for heteroepitaxial
  diffusion from ab initio data,'' \emph{Computational Materials Science},
  vol.~47, no.~2, pp. 577--583, 2009.

\bibitem[Vita and Trinkle(2024)]{vita2024spline}
J.~A. Vita and D.~R. Trinkle, ``Spline-based neural network interatomic
  potentials: Blending classical and machine learning models,''
  \emph{Computational Materials Science}, vol. 232, p. 112655, 2024.

\bibitem[Morawietz and Artrith(2021)]{morawietz2021machine}
T.~Morawietz and N.~Artrith, ``Machine learning-accelerated quantum
  mechanics-based atomistic simulations for industrial applications,''
  \emph{Journal of Computer-Aided Molecular Design}, vol.~35, no.~4, pp.
  557--586, 2021.

\bibitem[Zuo et~al.(2020)Zuo, Chen, Li, Deng, Chen, Behler, Cs{\'a}nyi,
  Shapeev, Thompson, Wood, et~al.]{zuo2020performance}
Y.~Zuo, C.~Chen, X.~Li, Z.~Deng, Y.~Chen, J.~Behler, G.~Cs{\'a}nyi, A.~V.
  Shapeev, A.~P. Thompson, M.~A. Wood \emph{et~al.}, ``Performance and cost
  assessment of machine learning interatomic potentials,'' \emph{The Journal of
  Physical Chemistry A}, vol. 124, no.~4, pp. 731--745, 2020.

\bibitem[Deringer et~al.(2019)Deringer, Caro, and
  Cs{\'a}nyi]{deringer2019machine}
V.~L. Deringer, M.~A. Caro, and G.~Cs{\'a}nyi, ``Machine learning interatomic
  potentials as emerging tools for materials science,'' \emph{Advanced
  Materials}, vol.~31, no.~46, p. 1902765, 2019.

\bibitem[Behler(2016)]{behler2016perspective}
J.~Behler, ``Perspective: Machine learning potentials for atomistic
  simulations,'' \emph{The Journal of chemical physics}, vol. 145, no.~17,
  2016.

\bibitem[Kobayashi et~al.(2017)Kobayashi, Giofr{\'e}, Junge, Ceriotti, and
  Curtin]{kobayashi2017neural}
R.~Kobayashi, D.~Giofr{\'e}, T.~Junge, M.~Ceriotti, and W.~A. Curtin, ``Neural
  network potential for al-mg-si alloys,'' \emph{Physical Review Materials},
  vol.~1, no.~5, p. 053604, 2017.

\bibitem[Dickel et~al.(2021)Dickel, Nitol, and Barrett]{dickel2021lammps}
D.~Dickel, M.~Nitol, and C.~Barrett, ``Lammps implementation of rapid
  artificial neural network derived interatomic potentials,''
  \emph{Computational Materials Science}, vol. 196, p. 110481, 2021.

\bibitem[Mishin(2021)]{mishin2021machine}
Y.~Mishin, ``Machine-learning interatomic potentials for materials science,''
  \emph{Acta Materialia}, vol. 214, p. 116980, 2021.

\bibitem[Shapeev(2016)]{shapeev2016moment}
A.~V. Shapeev, ``Moment tensor potentials: A class of systematically improvable
  interatomic potentials,'' \emph{Multiscale Modeling \& Simulation}, vol.~14,
  no.~3, pp. 1153--1173, 2016.

\bibitem[Bart{\'o}k et~al.(2013)Bart{\'o}k, Kondor, and
  Cs{\'a}nyi]{bartok2013representing}
A.~P. Bart{\'o}k, R.~Kondor, and G.~Cs{\'a}nyi, ``On representing chemical
  environments,'' \emph{Physical Review B—Condensed Matter and Materials
  Physics}, vol.~87, no.~18, p. 184115, 2013.

\bibitem[Gubaev et~al.(2019)Gubaev, Podryabinkin, Hart, and
  Shapeev]{gubaev2019accelerating}
K.~Gubaev, E.~V. Podryabinkin, G.~L. Hart, and A.~V. Shapeev, ``Accelerating
  high-throughput searches for new alloys with active learning of interatomic
  potentials,'' \emph{Computational Materials Science}, vol. 156, pp. 148--156,
  2019.

\bibitem[Podryabinkin et~al.(2019)Podryabinkin, Tikhonov, Shapeev, and
  Oganov]{podryabinkin2019accelerating}
E.~V. Podryabinkin, E.~V. Tikhonov, A.~V. Shapeev, and A.~R. Oganov,
  ``Accelerating crystal structure prediction by machine-learning interatomic
  potentials with active learning,'' \emph{Physical Review B}, vol.~99, no.~6,
  p. 064114, 2019.

\bibitem[Novikov et~al.(2018)Novikov, Suleimanov, and
  Shapeev]{novikov2018automated}
I.~S. Novikov, Y.~V. Suleimanov, and A.~V. Shapeev, ``Automated calculation of
  thermal rate coefficients using ring polymer molecular dynamics and
  machine-learning interatomic potentials with active learning,''
  \emph{Physical Chemistry Chemical Physics}, vol.~20, no.~46, pp.
  29\,503--29\,512, 2018.

\bibitem[Nitol et~al.(2025)Nitol, Mishra, Xu, and Fensin]{nitol2025moment}
M.~S. Nitol, A.~Mishra, S.~Xu, and S.~J. Fensin, ``Moment tensor potential and
  its application in the ti-al-v multicomponent system,'' \emph{Physical Review
  Materials}, vol.~9, no.~6, p. 063601, 2025.

\bibitem[Hafner(2008)]{hafner2008ab}
J.~Hafner, ``Ab-initio simulations of materials using vasp: Density-functional
  theory and beyond,'' \emph{Journal of computational chemistry}, vol.~29,
  no.~13, pp. 2044--2078, 2008.

\bibitem[Perdew et~al.(1996)Perdew, Burke, and
  Ernzerhof]{perdew1996generalized}
J.~P. Perdew, K.~Burke, and M.~Ernzerhof, ``Generalized gradient approximation
  made simple,'' \emph{Physical review letters}, vol.~77, no.~18, p. 3865,
  1996.

\bibitem[Nitol et~al.(2023)Nitol, Dang, Fensin, Baskes, Dickel, and
  Barrett]{nitol2023hybrid}
M.~S. Nitol, K.~Dang, S.~J. Fensin, M.~I. Baskes, D.~E. Dickel, and C.~D.
  Barrett, ``Hybrid interatomic potential for sn,'' \emph{Physical Review
  Materials}, vol.~7, no.~4, p. 043601, 2023.

\bibitem[Thompson et~al.(2022)Thompson, Aktulga, Berger, Bolintineanu, Brown,
  Crozier, In't~Veld, Kohlmeyer, Moore, Nguyen, et~al.]{thompson2022lammps}
A.~P. Thompson, H.~M. Aktulga, R.~Berger, D.~S. Bolintineanu, W.~M. Brown,
  P.~S. Crozier, P.~J. In't~Veld, A.~Kohlmeyer, S.~G. Moore, T.~D. Nguyen
  \emph{et~al.}, ``Lammps-a flexible simulation tool for particle-based
  materials modeling at the atomic, meso, and continuum scales,''
  \emph{Computer Physics Communications}, vol. 271, p. 108171, 2022.

\bibitem[Stukowski(2009)]{stukowski2009visualization}
A.~Stukowski, ``Visualization and analysis of atomistic simulation data with
  ovito--the open visualization tool,'' \emph{Modelling and simulation in
  materials science and engineering}, vol.~18, no.~1, p. 015012, 2009.

\bibitem[Nitol et~al.(2022)Nitol, Mun, Dickel, and
  Barrett]{nitol2022unraveling}
M.~S. Nitol, S.~Mun, D.~E. Dickel, and C.~D. Barrett, ``Unraveling mg< c+ a>
  slip using neural network potential,'' \emph{Philosophical Magazine}, vol.
  102, no.~8, pp. 651--673, 2022.

\bibitem[Warlimont and Martienssen(2018)]{warlimont2018springer}
H.~Warlimont and W.~Martienssen, \emph{Springer handbook of materials
  data}.\hskip 1em plus 0.5em minus 0.4em\relax Springer, 2018.

\bibitem[Su et~al.(2019)Su, Xu, and Beyerlein]{su2019density}
Y.~Su, S.~Xu, and I.~J. Beyerlein, ``Density functional theory calculations of
  generalized stacking fault energy surfaces for eight face-centered cubic
  transition metals,'' \emph{Journal of Applied Physics}, vol. 126, no.~10,
  2019.

\bibitem[Kittel and McEuen(2018)]{kittel2018introduction}
C.~Kittel and P.~McEuen, \emph{Introduction to solid state physics}.\hskip 1em
  plus 0.5em minus 0.4em\relax John Wiley \& Sons, 2018.

\bibitem[Siegel(1978)]{siegel1978vacancy}
R.~Siegel, ``Vacancy concentrations in metals,'' \emph{Journal of Nuclear
  Materials}, vol.~69, pp. 117--146, 1978.

\bibitem[Liyanage et~al.(2024)Liyanage, Turlo, and Curtin]{liyanage2024machine}
M.~Liyanage, V.~Turlo, and W.~Curtin, ``Machine learning potential for the cu-w
  system,'' \emph{Physical Review Materials}, vol.~8, no.~11, p. 113804, 2024.

\bibitem[Andolina et~al.(2021)Andolina, Bon, Passerone, and
  Saidi]{andolina2021robust}
C.~M. Andolina, M.~Bon, D.~Passerone, and W.~A. Saidi, ``Robust,
  multi-length-scale, machine learning potential for ag--au bimetallic alloys
  from clusters to bulk materials,'' \emph{The Journal of Physical Chemistry
  C}, vol. 125, no.~31, pp. 17\,438--17\,447, 2021.

\bibitem[Hunter et~al.(2013)Hunter, Zhang, Beyerlein, Germann, and
  Koslowski]{hunter2013dependence}
A.~Hunter, R.~Zhang, I.~J. Beyerlein, T.~C. Germann, and M.~Koslowski,
  ``Dependence of equilibrium stacking fault width in fcc metals on the
  $\gamma$-surface,'' \emph{Modelling and Simulation in Materials Science and
  Engineering}, vol.~21, no.~2, p. 025015, 2013.

\bibitem[Touloukian(1975)]{touloukian1975thermal}
\BIBentryALTinterwordspacing
Y.~Touloukian, \emph{Thermal Expansion: Metallic Elements and Alloys}, ser.
  TPRC data series.\hskip 1em plus 0.5em minus 0.4em\relax Springer US, 1975.
  [Online]. Available: \url{https://books.google.com/books?id=XR1NAQAAIAAJ}
\BIBentrySTDinterwordspacing

\bibitem[Baty et~al.(2021{\natexlab{a}})Baty, Burakovsky, and
  Errandonea]{baty2021cu}
S.~R. Baty, L.~Burakovsky, and D.~Errandonea, ``Ab initio phase diagram of
  copper,'' \emph{Crystals}, vol.~11, no.~5, p. 537, 2021.

\bibitem[Baty et~al.(2021{\natexlab{b}})Baty, Burakovsky, and
  Errandonea]{baty2021ag}
S.~Baty, L.~Burakovsky, and D.~Errandonea, ``Ab initio phase diagram of
  silver,'' \emph{Journal of Physics: Condensed Matter}, vol.~33, no.~48, p.
  485901, 2021.

\bibitem[Sadigh et~al.(2012)Sadigh, Erhart, Stukowski, Caro, Martinez, and
  Zepeda-Ruiz]{sadigh2012scalable}
B.~Sadigh, P.~Erhart, A.~Stukowski, A.~Caro, E.~Martinez, and L.~Zepeda-Ruiz,
  ``Scalable parallel monte carlo algorithm for atomistic simulations of
  precipitation in alloys,'' \emph{Physical Review B—Condensed Matter and
  Materials Physics}, vol.~85, no.~18, p. 184203, 2012.

\bibitem[Metropolis et~al.(1953)Metropolis, Rosenbluth, Rosenbluth, Teller, and
  Teller]{metropolis1953equation}
N.~Metropolis, A.~W. Rosenbluth, M.~N. Rosenbluth, A.~H. Teller, and E.~Teller,
  ``Equation of state calculations by fast computing machines,'' \emph{The
  journal of chemical physics}, vol.~21, no.~6, pp. 1087--1092, 1953.

\end{thebibliography}

\end{document}